# Performance characteristics of low threshold current 1.25-µm type-II GaInAs/GaAsSb "W"-lasers for optical communications

Dominic A. Duffy[1], Igor P. Marko[1], Christian Fuchs[2], Timothy D. Eales[1], Jannik Lehr[2], Wolfgang Stolz[2], and Stephen J. Sweeney[1*]

[1] *Advanced Technology Institute and Department of Physics, University of Surrey, Guildford, GU2 7XH, United Kingdom*

[2] *Materials Sciences Centre and Department of Physics, Philipps-Universität Marburg, Renthof 5, 35032, Marburg, Germany*

**email: s.sweeney@surrey.ac.uk*

## Abstract

Type-II "W"-lasers have made an important contribution to the development of mid-infrared laser diodes. In this paper, we show that a similar approach can yield high performance lasers in the optical communications wavelength range. (GaIn)As/Ga(AsSb) type-II "W" structures emitting at 1255nm have been realised on a GaAs substrate and exhibit low room temperature threshold current densities of 200-300 A cm$^{-2}$, pulsed output powers exceeding 1 W for 100µm wide stripes, and a characteristic temperature T$_0$ ≈90K around room temperature. Optical gain studies indicate a high modal gain around 15-23 cm$^{-1}$ at 200-300 A cm$^{-2}$ and low optical losses of 8±3 cm$^{-1}$. Analysis of the spontaneous emission indicates that at room temperature, up to 24% of the threshold current is due to radiative recombination, with the remaining current due to other thermally activated non-radiative processes. The observed decrease in differential quantum efficiency with increasing temperature suggests that this is primarily due to a carrier leakage process. The impact of these processes is discussed in terms of the potential for further device optimisation. Our results present strong figures of merit for near-infrared type-II laser diodes and indicate significant potential for their applications in optical communications.

## Introduction

The development of increased efficiency and more temperature stable semiconductor lasers in the near-infrared (NIR) wavelength range is important for the future of lower energy-consuming data communications networks. This has inspired a range of approaches to achieve more temperature stable operation including the use of new semiconductor materials in type-I inter-band quantum well (QW) laser systems such as (AlGaIn)As/InP [1], [2], (GaIn)(NAs)/GaAs [3]–[5], Ga(AsBi)/GaAs [6], [7], [8] and the use of quantum dot-based active regions [9], [10]. While these approaches have demonstrated considerable progress, devices remain limited by non-radiative recombination processes such as Auger recombination, carrier leakage, or defect-related recombination associated with less mature material growth [11], [12], [13], [7]. Of these, Auger recombination is a fundamental process, which may be reduced/suppressed through structural design (such as via type-II band alignments [14], [15]) or band structure engineering of the quantum well material e.g. using dilute bismides [6], [7], [13]. Carrier leakage processes are sensitive to the band offsets of the active region while defect-related recombination is sensitive to material quality, which can potentially be improved with growth and fabrication optimisation; this is a particular issue for new alloys [16]. In the mid-infrared, type-II so-called "W"-structures have shown great promise, providing a flexible approach to wavelength engineering [17], and potential as a route to reducing Auger recombination [14], [15]. This approach has recently been demonstrated in the near-infrared with (GaIn)As/Ga(AsSb) type-II structures which may be grown on GaAs [18]. The growth on GaAs substrates is attractive as it lends itself to the realisation of low-cost Vertical Cavity Surface Emitting Lasers (VCSELs) operating at longer wavelengths.

In type-II quantum well heterostructures (QWHs), electrons and holes are spatially separated in adjacent layers [19]. As a result of this separation, the electron and hole behaviours are mainly determined by different materials, thus allowing for independent tuning of the conduction and valence band properties in the active region. However, this separation also results in a reduced wave function overlap between electron and hole states when compared to type-I structures, which can reduce the probability of radiative recombination. "W"-quantum well heterostructures ("W"-QWHs) can overcome this issue by sandwiching a thin hole quantum well between two electron quantum wells, resulting in a "W"-shaped conduction band confinement potential. This leads to an increased electron-hole wave function overlap while still preserving the 2D density of states and providing greater control over the carrier properties [20] (as illustrated in Fig. 1).

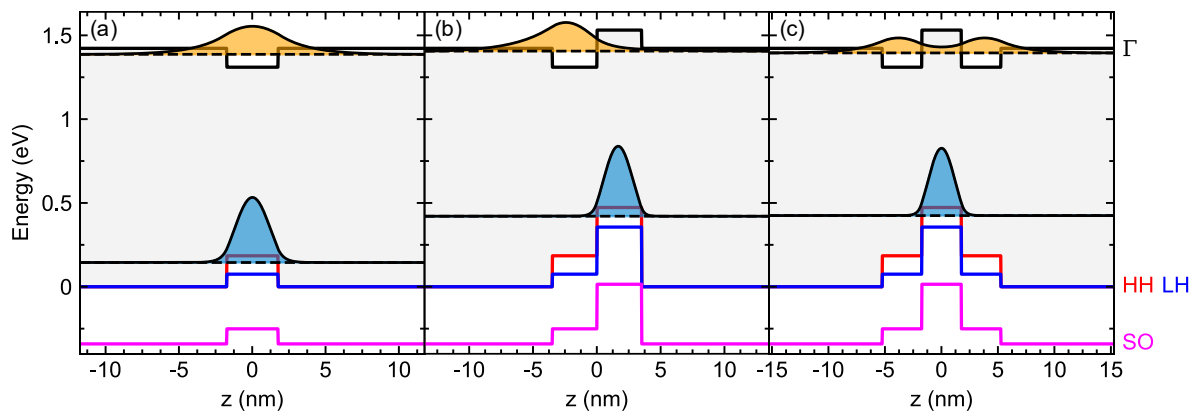

*Figure 1. Illustration of the active region of a single (a) type-I (b) type-II (c) type-II "W"-QW. In the "W"-QW, electrons are mostly confined in the (GaIn)As quantum wells, with holes confined in the central Ga(AsSb) quantum well. The Ga(AsSb) quantum well is thin enough for the electrons to tunnel, resulting in an increased wavefunction overlap with the hole states compared to a standard type-II design.*

A combination of quantum well materials consisting of a Ga(AsSb) hole-confining quantum well surrounded by (GaIn)As electron-confining quantum wells allows for the "W"-QWH to be grown pseudomorphically on GaAs substrates. While these heterostructures are promising candidates for designing efficient active regions, previous (GaIn)As/Ga(AsSb)-based devices have been demonstrated at wavelengths which were either too short for standard 1.3 μm telecommunication applications e.g. at 1164 nm [21], had a tendency to switch to higher-order (and/or type-I) transitions under increasing injection [22], or exhibited relatively large threshold current densities (> 1000 A cm$^{-2}$) [18], [23]. Much of the physics underpinning the current understanding of type-II structures is not well-established, with many theoretical models based upon theory developed for simpler type-I structures which may not always be applicable to type-II structures [14]. Thus, in order to optimise the performance of these structures an understanding of the characteristics and main optical and electronic processes occurring in these "W"-QWH devices operating in the near-infrared is essential.

In this paper, we report on the device performance of (GaIn)As/Ga(AsSb)-based lasers operating at 1255 nm as a function of temperature and current, comparing and contrasting this behaviour with standard type-I lasers emitting at similar wavelengths. The investigated device structure is based on a double "W"-QWH design [18], with improvements to both the epitaxial process and structural design. Improvements in growth are achieved by scaling the epitaxy process to full 2-inch wafers for improved homogeneity across the sample as well as by growing the devices in a single epitaxy run for improved material quality. The device structure is based upon previously demonstrated designs [18] but with an increased thickness of the (AlGa)As cladding layers to improve optical confinement.

In this study, to support the optimisation of lasers using this "W"-QWH approach, we measure the optical gain and absorption spectra of segmented contact devices and so determine the internal optical losses in these structures. Measurements of the temperature dependence of the threshold current as well as pure spontaneous emission are then used to quantify the role of radiative and non-radiative recombination mechanisms in these devices and their influence on device performance.

## Device Structures

The laser heterostructures used in this work are grown by metalorganic vapour-phase epitaxy on full 2-inch n-doped GaAs (001) substrates, using a growth procedure similar to that described by Fuchs *et al.* [18]. To grow these structures, a commercial AIXTRON AIX 200-GFR reactor system with a carrier gas of $H_2$ is used. During growth, triethylgallium (TEGa) and trimethylindium (TMIn) are used as group-III precursors, with triethylantimony (TESb) and tertiarybutylarsine (TBAs) used as group-V precursors. After treating the substrate with a tertiarybutylarsine-stabilised bake-out procedure to remove the native oxide layer, epitaxial growth is carried out at a temperature of 550°C for the active regions and 625°C for the surrounding layers.

During deposition, a 0.2 µm thick n-GaAs buffer is first added to the substrate in order to obtain a high-quality growth surface. This buffer is followed by a 1.5 µm n-$(Al_{0.36}Ga_{0.64})$As cladding layer and a 0.2 µm undoped GaAs separate confinement heterostructure layer. The active region of the structure is then grown by depositing two "W"-QWHs separated by a 20 nm GaAs barrier, where each "W"-QWH consists of a 3.5 nm $Ga(As_{0.73}Sb_{0.27})$ hole quantum well sandwiched between two 3.5 nm $(Ga_{0.73}In_{0.27})$As electron quantum wells. Following the deposition of the active region, the p-side of the device once again consists of a 0.2 µm undoped GaAs separate confinement heterostructure layer and a 1.5 µm p-$(Al_{0.36}Ga_{0.64})$As cladding layer. Finally, a 0.2 µm highly doped p-GaAs cap is added by doping a GaAs layer using $CBr_4$ in order to ensure small contact resistances. The resulting layer stack is shown in Fig. 2(a).

Prior to device fabrication, the nominal compositions and thicknesses of the grown heterostructure are measured with a Panalytical X'Pert Pro diffractometer using standard high-resolution X-ray diffraction (HR-XRD) techniques [24] at the centre of the wafer. The resulting measurement and calculated HR-XRD fit (assuming uniform layers with abrupt interfaces) are shown in Fig. 2(b) for the (004) reflection, with the excellent agreement between theory and experiment alongside well-defined interference fringes observable up to -6000" indicating high structural quality.

The wafer is then processed as outlined in Ref [21] to obtain gain-guided broad-area edge-emitting lasers with a cavity length of approximately 1 mm and a contact width of 100 µm. To investigate the behaviour of radiative current in the devices, spontaneous emission (SE) spectra [25] are obtained from circular windows fabricated into the n-type contact using a focused ion beam.

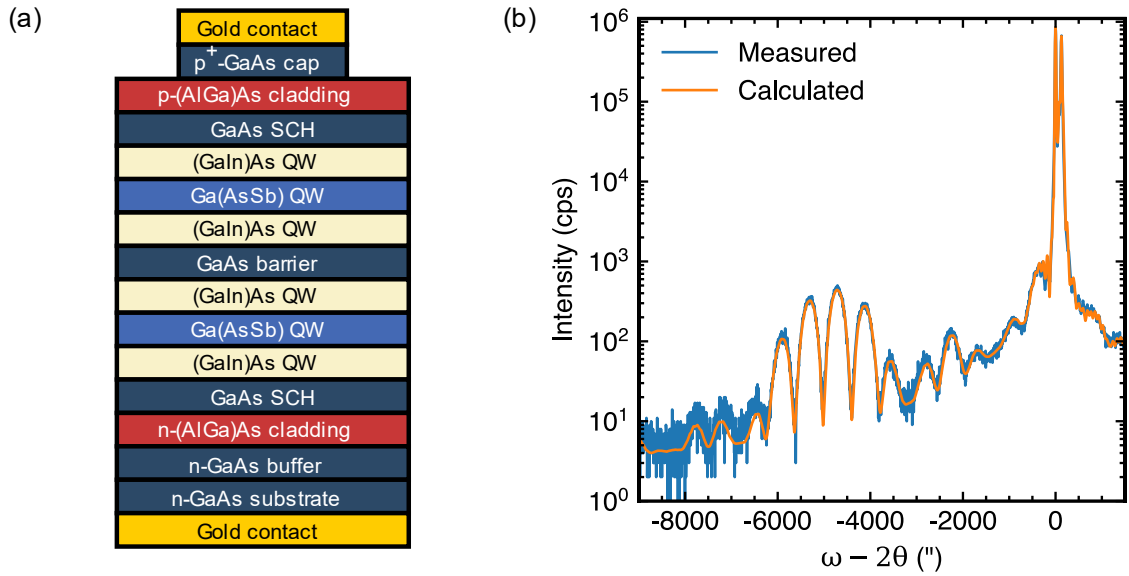

Figure 2. (a) Nominal device structure. (b) (004) HR-XRD measurement and simulated fit used to extract the compositions and thicknesses of the heterostructure.

## Optical Gain and Loss Measurements

Experimental measurements of optical gain and absorption in these devices are performed using the segmented contact technique developed by Blood *et al.* [26]. Top stripe contacts are fabricated using a mask to produce segments 100 μm wide and 250 μm in length separated by 3 μm spacers [27]. To form the metal contacts, Ti/Au-based contacts are deposited onto both the top of the p-GaAs contact layer and the underside of the n-GaAs substrate. The samples are then chemically etched to remove the uncontacted p-doped top layer of the structures in order to avoid electrical connection between adjacent segments. To avoid round trip light amplification long pieces of material 10-20 mm away from the wafer edge are used, providing long passive regions behind the segments to act as absorbers with the non-cleaved wafer edge used as an additional measure to suppress optical feedback.

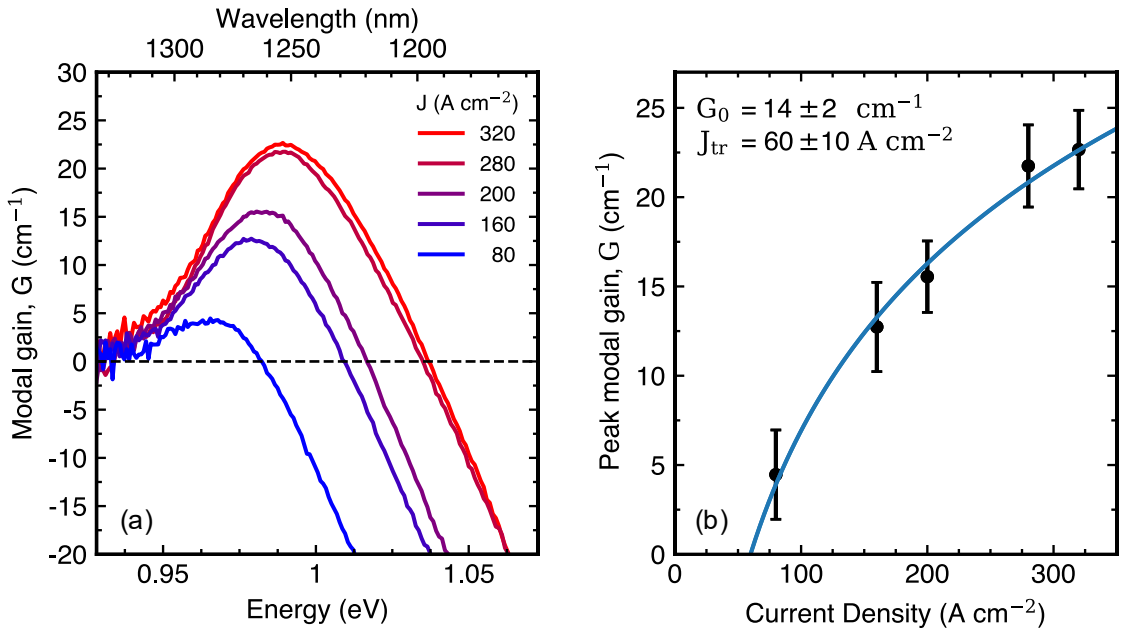

*Figure 3. (a) Room-temperature net modal gain spectra at varying current densities in range 80-320 A cm$^{-2}$ (b) peak modal gain as a function of current density. The blue line shows the fit of a simplified gain-current curve for a single sub-band pair, from which the transparency current can be extracted* [28].

Using this approach, a set of room temperature (20°C) net modal gain (G-$\alpha_i$) and net modal absorption (A+$\alpha_i$) measurements are obtained, where $\alpha_i$ represents the internal optical losses (e.g. through scattering, free carrier, and inter-valence band absorption). Over the range of current densities measured, 80-320 A cm$^{-2}$, a typical optical loss of $\alpha_i$ = 8±3 cm$^{-1}$ is observed which is comparable to that of type-I devices operating at similar wavelengths [29]. The measured optical losses can then be used to calculate the modal gain (G) as shown in Fig. 3(a). From this, the peak modal gain as a function of current density is extracted and plotted in Fig. 3(b), with peak modal gains of up to G ≈23 cm$^{-1}$ measured. By fitting a simplified gain-current relation model for a single sub band transition [28] the current density of transparency is estimated as J$_{tr}$ = 60±10 A cm$^{-2}$ with a modal material gain of $G_0$ = 14±2 cm$^{-1}$. At current densities approaching threshold in equivalent laser structures, the gain peak reaches a full width at half maximum (FWHM) of ≈55 meV and a peak position of ≈988 meV (≈1255 nm).

These values for peak gain and FWHM compare favourably to previous work on W-structures targeting similar wavelengths, where current densities up to J ≈4.0 kA/cm$^2$ were required to achieve comparable peak gains [23], and multiple transitions can be observed contributing to a broader gain spectrum with a FWHM of ≈90 meV. These improvements are attributed to the improved design and growth quality of the current structure. However, compared to standard type-I (GaIn)(AsP)/InP devices operating at similar wavelengths, with FWHMs of ≈31 meV [30], the gain spectra of these "W"-QWHs is still broader, which we attribute to a combination of type-II related band filling effects and compositional fluctuations (e.g. in the Ga(AsSb) well) which result in inhomogeneous broadening. Additionally, we note a significant blue shift of the gain peak with increasing current density, with a shift of ΔE ≈23 meV going from 80 to 280 A cm$^{-2}$ compared to ≈6 meV in type-I (GaIn)(AsP)/InP over a similar fraction of threshold [30]. This is attributed to the charge separation effects inherent to the type-II "W"-QWHs investigated in this work, which gives rise to enhanced band bending effects with increasing carrier density.

## Temperature-Dependent Characteristics (I) Lasing

Among the main requirements for semiconductor lasers are their energy efficiency and temperature stability. Therefore, their threshold current density, $J_{th}$, should be as low as possible with a low (ideally zero) temperature sensitivity. In practice, an increasing exponential dependence of $J_{th}$ versus temperature is typically observed over a particular temperature range, characterised by a characteristic temperature, $T_0$, which is also a function of temperature [31]. This degradation in performance with increasing temperature is often due to the impact of non-radiative recombination [16] and carrier leakage [32] processes and is sometimes affected by temperature-dependent optical losses [33]. These recombination and loss mechanisms have different dependencies on carrier density, $n$, and temperature, $T$. Hence, temperature-dependent measurements of $J_{th}$ can provide a useful insight into the dominant processes contributing to the device characteristics.

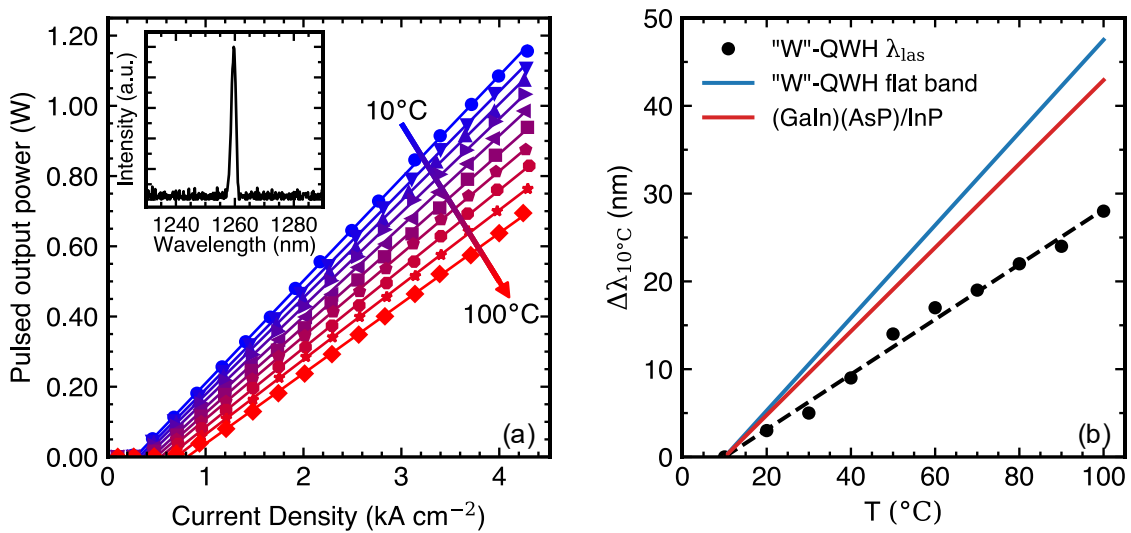

*Figure 4. (a) power-current (L-I) characteristics for heatsink temperatures between 10°C and 100°C in steps of 10°C, with a spectral measurement above threshold at 20°C included in the inset showing lasing at 1255 nm, (b) Measured variation in lasing wavelength with temperature relative to its value at 10°C for the measured "W"-QWHs (black points and dashed line) compared against flat-band calculations of the "W"-QWH (blue line) and a type-I (GaIn)(AsP)/InP device (red line) [34].*

At elevated temperatures, calibrated power measurements are obtained on a probe station over a range of 10-100°C with emission detected using a large area germanium detector. Fig. 4(a) shows the measured pulsed output power versus current density characteristics (light-current/L-I curves) over this range with lasing achieved up to the maximum operating temperature and power outputs exceeding 0.7 W at 100°C. The measurement at 20°C is carried out before and after the temperature series to verify that no device degradation occurs.

The inset in Fig. 4(a) shows the lasing spectrum measured above threshold at 20°C, with the lasing peak at 1255 nm in good agreement with the spectral position of the maximum gain peak at similar current densities. The room temperature threshold current density was found to be in the range of 200-300 A cm$^{-2}$ with some scatter between laser chips suggesting a degree of non-uniformity across the wafer, while at room temperature the characteristic temperature $T_0$ is found to be ≈90K. We note that, despite the reduced electron-hole wavefunction overlap expected for type-II active regions, these devices show low $J_{th}$ values comparable to good type-I devices operating around this wavelength [35], [36] along with slightly higher room temperature $T_0$ values compared to $T_0 ≈60$K observed in type-I devices suggesting a more thermally stable device.

Additionally, the measured lasing wavelength at threshold over this temperature range is plotted in Fig. 4(b), and shows a reduced thermal red shift of $\frac{d\lambda}{dT} = 0.31 \pm 0.01$ nm/°C when compared to both experimental measurements of typical type-I InGaAsP/InP devices ($\frac{d\lambda}{dT} \approx 0.4 - 0.48$ nm/°C [34]) and calculations of the type-II transition energy in the investigated structure under flat band conditions ($\frac{d\lambda}{dT} = 0.53 \pm 0.01$ nm/°C [37]) over the same temperature range. This is likely due to the aforementioned charge separation effects in type-II structures and highlights the potential of utilising these effects to improve the temperature stability of the wavelength of NIR lasers through careful device design.

To explore the temperature-dependent device behaviour over a wider temperature range, L-I measurements are collected using a closed-cycle helium cryostat system over the temperature range of 20-295K. To eliminate Joule heating effects, current was supplied in pulsed mode with a 0.05% duty cycle (1 kHz, 500 ns pulse width). The emitted light is collected with a (GaIn)As integrating sphere. Measurements of the temperature dependence of threshold current density $J_{th}$ are presented in Fig. 5 (open black squares) along with measurements of the radiative current density $J_{rad}$ (open blue circles) obtained from the integrated spontaneous emission (SE) spectra $L_{spon}$ at threshold.

To normalise $J_{rad}$ for comparison to $J_{th}$ we assume that at low temperature non-radiative processes are negligible, and so the low-temperature threshold current should be dominated by radiative processes only (i.e. $J_{th} = J_{rad}$ at low T). This assumption is supported by the observation that $J_{rad}$ closely matches $J_{th}$ up to ≈110K; we also observe low absolute $J_{th}$ values and an absence of a super-linear relation between integrated SE vs current, which would be a characteristic feature of defect-related recombination. From these measurements, the relative contributions of (maximum) radiative and (minimum) non-radiative recombination processes to $J_{th}$ can be extracted, as shown by the blue ($J_{rad}$) and red ($J_{nonrad}$) hatched regions of Fig. 5.

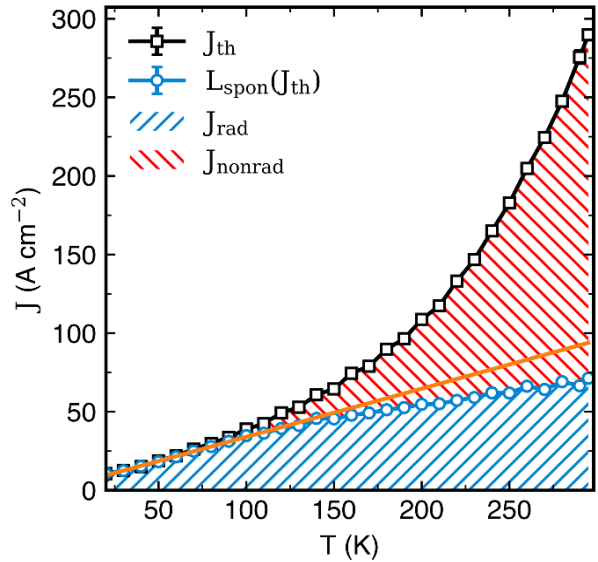

*Figure 5. The temperature dependence of threshold current density $J_{th}$ (open black squares) and normalised integrated spontaneous emission $L_{spon}$ at $J_{th}$ (open blue circles), along with the estimated radiative ($J_{rad}$, blue hatched region) and non-radiative ($J_{nonrad}$, red hatched region) current densities of an as-cleaved Fabry-Perot laser diode. The orange line is a guide for the eye highlighting the deviation from the expected linear radiative behaviour with increasing temperature.*

From Fig. 5, it can be seen that the temperature dependence of $J_{th}$ is approximately linear at low temperature with a super-linear variation towards higher temperatures, similar to that observed in type-I devices [1]. Meanwhile, the temperature dependence of $J_{rad}$ is similarly linear at low temperatures up to 110-120K with an unusual sub-linear variation towards higher temperatures (for contrast, the orange line in Fig. 5 shows the extrapolated linear fit to $J_{rad}$ at low T). As a result of this sub-linear $J_{rad}$, we estimate that the radiative current constitutes no more than 24±1% ($J_{rad} \approx$ 71 A cm$^{-2}$) of $J_{th}$ at room temperature with non-radiative processes accounting for the remaining 76±1% ($J_{nonrad} \approx$ 219 A cm$^{-2}$). If the behaviour of $J_{rad}$ were linear, as indicated by the orange line in Fig. 5, this would still result in a highly non-radiatively dominated device with 32±3% of $J_{th}$ from $J_{rad}$ and 68±3% $J_{nonrad}$. While the relative fraction of radiative current in these devices is lower than the approximately 50:50 split observed in 1.3 μm type-I (GaIn)(AsP)/InP devices [29] (which may be expected due to the reduced wavefunction overlap in type-II active regions), on an absolute scale the overall $J_{th}$ in these devices is still low with absolute $J_{nonrad}$ contributions comparable to type-I devices.

This behaviour can be investigated further through a simple temperature dependence model for the radiative and non-radiative components of the threshold current density [38]. Assuming equal electron and hole densities (*n = p*) in the quantum wells, the current density flowing through the laser may be expressed as

$$J_{th} = eL(An + Bn^2 + Cn^3) + J_{leak}, \qquad (1)$$

where *e* is the electronic charge, *L* is the active region thickness, $J_{leak}$ is the carrier and temperature-dependent leakage current, and the three terms account for monomolecular (defect-related) recombination ($An$), radiative recombination ($Bn^2$), and Auger recombination ($Cn^3$), respectively. Assuming the recombination coefficients are independent of carrier concentration, for an ideal quantum well, the temperature dependence of the radiative and Auger coefficients can be expressed as [39]

$$B \propto T^{-1}, \qquad (2)$$

and

$$C = C_0 \exp(-E_a/k_B T), \qquad (3)$$

where we assume that the Auger process is thermally activated [40] with an activation energy, $E_a$, and where $T$ is the temperature and $k_B$ is the Boltzmann constant. For an ideal quantum well the carrier density at threshold, $n_{th} \propto T$. Hence, it follows that the radiative current, $J_{rad} \propto T$ [29]. We note that since the electrons and holes are spatially delocalized in type-II structures, the assumption that *n=p* may not be valid. Furthermore, *B* may itself be injection (*n*) dependent. From Fig. 6 we observe that the low temperature $J_{th}(T)$ and $J_{rad}(T)$ are linear, as expected from the simple model, suggesting that in this temperature range, the radiative current dominates in a manner consistent with an ideal QW laser. However, the increasingly sub-linear $J_{rad}$ with increasing temperature suggests that a more complex physical mechanism is occurring beyond the low-temperature regime. Several possible mechanisms could explain this behaviour, including phase-space filling [41] and injection-dependent radiative coefficients due to the spatial delocalisation of carriers in type-II structures. The details of this are beyond the scope of the present study, however additional theoretical work is underway to more conclusively identify the dominant process(es) leading to the observed $J_{rad}$ behaviour.

## Temperature-Dependent Characteristics (II) Spontaneous Emission Analysis

To help understand the origins of the unusual thermal behaviour of $J_{rad}$ we can examine the measured SE and resulting integrated SE, $L_{spon}$, used to calculate $J_{rad}$ in more detail. In an ideal radiatively-dominated device, $L_{spon}$ should increase linearly with current density up to $J_{th}$, after which it should flatten out as any additional carriers injected into the device recombine via stimulated emission – this results in a "pinning" of carrier concentration beyond threshold.

Fig. 6 shows $L_{spon}$ up to and beyond threshold at 70K (blue line), 200K (orange line), and 295K (green line). The threshold current density $J_{th}$ and resulting radiative current density at threshold $J_{rad}$ for each temperature are indicated by dashed vertical and horizontal lines, respectively. Additionally, the black dashed lines for 70K and 295K show the expected $L_{spon}(J)$ relation for an ideal radiatively-dominated device with full carrier pinning above threshold.

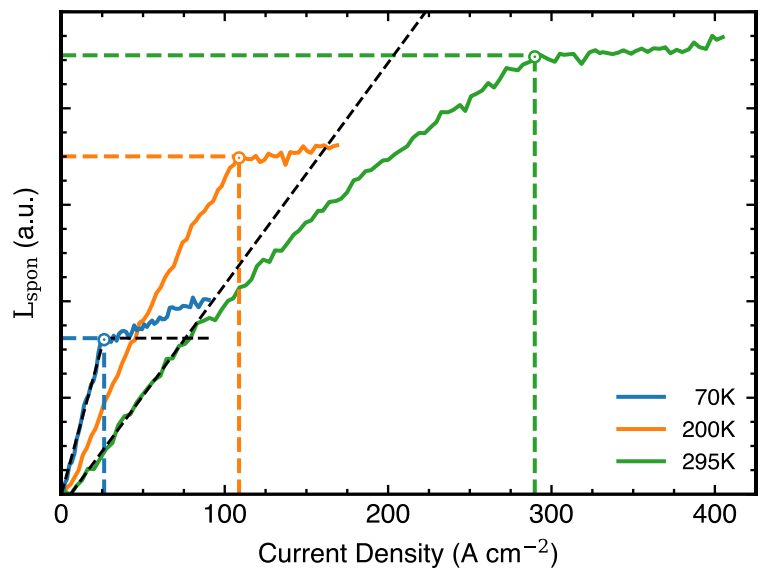

*Figure 6. Integrated spontaneous emission $L_{spon}$ as a function of current density for 70K, 200K, and 295K. Coloured dashed lines indicated the L-I determined threshold value and corresponding $L_{spon}$ used in Fig. 6. Black dashed lines overlaid on 70K and 295K highlight the expected $L_{spon}$ behaviour for an homogenous radiatively-dominated device.*

At low temperature, a linear $L_{spon}(J)$ can be observed up to threshold, which is in good agreement with a radiatively-dominated device. This further supports the validity of the assumption that $J_{rad} \approx J_{th}$ at low temperatures, which was previously used to normalise $J_{rad}$ in Fig. 5. With increasing temperature, however, an increasingly sub-linear $L_{spon}(J)$ is observed. This strongly indicates the presence of a thermally activated carrier density-dependent loss process that degrades device performance (such as Auger recombination or carrier leakage).

Above threshold, the low temperature $L_{spon}$ measurements show a significant level of non-pinning that decreases with increasing temperature. This improvement in carrier pinning is likely attributable to the thermalization of localized states in the active region (possibly due to a degree of inhomogeneity in Sb incorporation). Qualitatively, the relative non-pinning gradient decreases exponentially with increasing temperature, with approximately an order of magnitude decrease (i.e. improvement in carrier pinning) observed going from 20-295K.

Another notable feature of the measured SE spectra is the shift in energy of the SE peak with current density and temperature. This shift in SE is comparable to both the previously noted shifts in peak modal gain with J (Fig. 3(a)) and the reduced temperature sensitivity of the lasing energy at threshold (Fig. 4(b)), but allows for investigating this behaviour over a range of temperatures and current densities. Fig. 7(a) shows the measurements of shift in peak SE from low current densities as a fraction of $J_{th}$ for devices operating at 70K (blue points), 140K (purple points), and 295K (red points). The peak shift at threshold increases with increasing temperature. This can be explained by the increase in current density at threshold $J_{th}$ with temperature leading to an increase in carrier density in the active region. In the case of a type-II active region, the spatially separated charge carriers result in significant band bending effects. This blue shift acts to partially compensate for the typical thermal redshift observed in semiconductor lasers due to band gap shrinkage (e.g. as described by the Varshni relation [42]), thus resulting in the overall decrease in temperature sensitivity of the lasing wavelength observed in Fig. 4(b). Beyond threshold, this shift plateaus due to carrier pinning – it is interesting to note that this contrasts with the clear non-pinning observed in low temperature $L_{spon}$ measurements (Fig. 6).

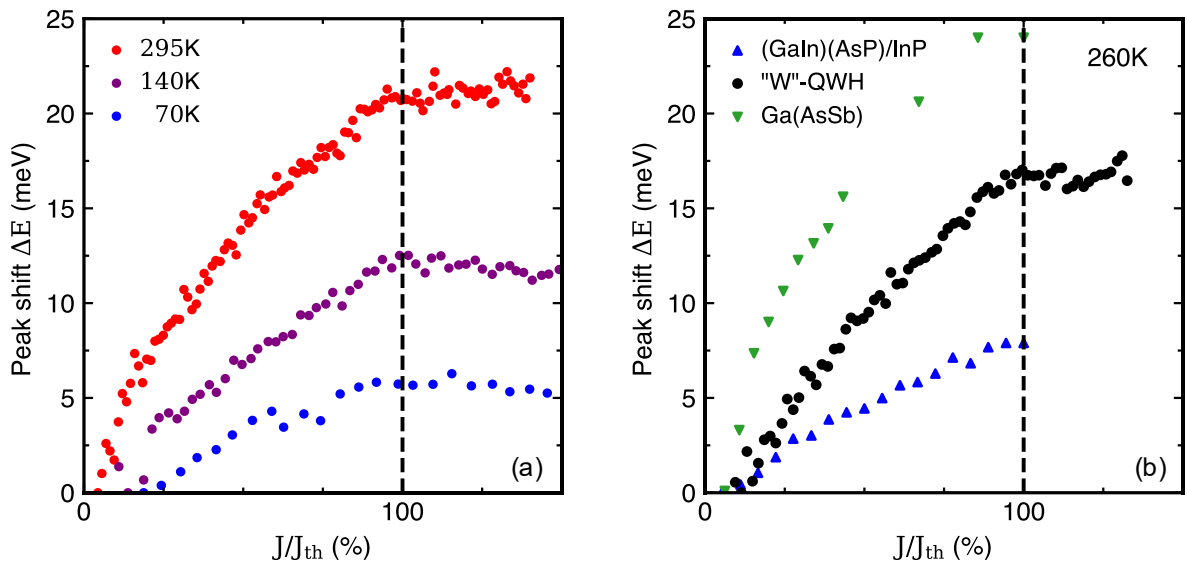

*Figure 7. SE peak shift from low current as a function of current density and normalised to threshold for (a) 70K, 140K, and 295K "W"-QWHs (b) "W"-QWHs compared to previously reported shifts from (GaIn)(AsP)/InP and Ga(AsSb)/GaAs devices at 260K [43] .*

Additionally, Fig. 7(b) shows the peak shift for the investigated device structure (black circles) compared to literature measurements of both a type-I (GaIn)(AsP)/InP device (blue triangles) and a type-II Ga(AsSb)/GaAs device (green triangles) recorded at a temperature of 260K [43]. While all devices exhibit an increasing blueshift with increasing current density, this blueshift is lowest for type-I systems with strong carrier localisation and enhanced for type-II alignments where carrier separation plays a more significant role. The intermediate shift observed in the measured "W"-QWH tentatively suggests the possibility of tuning device structures to modify the temperature dependence of the emission wavelength.

## Temperature-Dependent Characteristics (III) Differential Efficiency

To further investigate the origin of the non-radiative processes dominating the high-temperature device characteristics, we turn our attention to the temperature dependence of the differential efficiency, $\eta_d$, obtained from the linear gradient of the L-I measurements above threshold (i.e. where stimulated emission dominates). By normalising the low-temperature L-I data to the values around room temperature obtained from the initial high temperature calibrated L-I measurements, we obtain a consistent set of $\eta_d$ across a temperature range of 20-370K as shown in Fig. 8. We note that this normalisation gives good agreement in the continuity of the gradient between datasets, with the scatter at low-temperature likely due to the temperature stability of the cryostat.

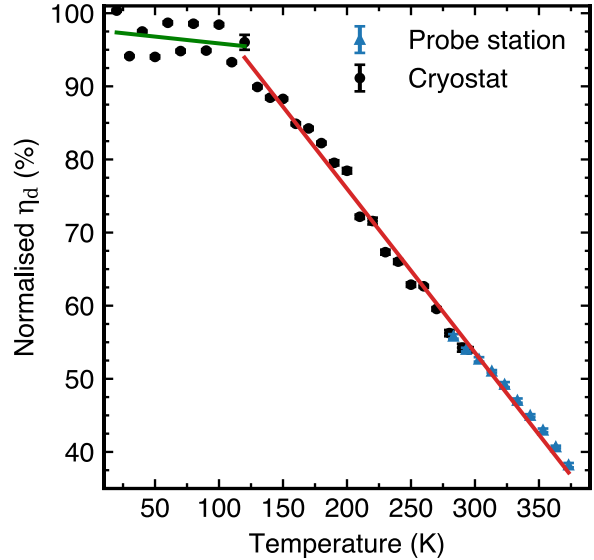

*Figure 8. Normalised differential efficiency $\eta_d$ as a function of temperature for low temperature cryostat L-Is (black circles) and high temperature probe station L-Is (blue triangles). Data is normalised to $\eta_d$ values obtained at 10-20ºC. Lines are included as guides to the eye showing regions of roughly constant (green) and linearly decreasing (red) $\eta_d$ with a transition between these two regimes occurring at ≈110-120K.*

From Fig. 8, $\eta_d$ is initially flat/plateaued up to around 110-120K, beyond which a consistent linear decrease in $\eta_d$ is observed with increasing temperature up to the maximum temperature measured. Interestingly, this "break-point"-like behaviour coincides with both the onset of sub-linear $J_{rad}$ and super-linear $J_{nonrad}$, shown in Fig. 5, and indicates the presence of carrier leakage or temperature-dependent optical losses. Given the modest optical losses observed in the modal gain measurements and the fact that $J_{rad}$ increases sub-linearly with temperature, temperature-dependent optical losses are unlikely to explain the significant variation in $\eta_d$ observed. On the other hand, thermal leakage of carriers from the active region is plausible due to both the low electron confinement of the designed structure (≈28-41 meV for carrier densities ranging from flat band conditions to transparency) and the delocalisation of carriers that are characteristic of a type-II active region. When combined with the previously noted indications of non-radiative recombination from analysis of SE and $J_{rad}/J_{nonrad}$, these results are consistent with thermally activated carrier leakage dominating the temperature-dependent properties of these devices.

It is also interesting to note the discrepancy between the decrease in SE non-pinning observed in Fig. 6 (which is expected to improve $\eta_d$) and the overall decrease in $\eta_d$ with temperature, which indicates that thermalisation of localised states has a smaller impact on $J_{th}$ than the thermally activated leakage processes. This may suggest that refinements to future device design (e.g. by improving electron confinement) would result in more significant improvements in overall device performance than optimisation of growth quality to reduce carrier localisation effects would by itself.

## Conclusions

In this study we have investigated the carrier recombination and temperature-dependent behaviour of type-II "W"-(GaIn)As/Ga(AsSb) lasers emitting at around 1255nm. Based upon segmented contact measurements we observe strong modal gain characteristics up to G ≈23 cm$^{-1}$ and relatively low optical losses of $\alpha_i$ =8±3 cm$^{-1}$ at room temperature. Broad area lasers processed from the wafer exhibit relatively low pulsed room temperature threshold current densities of ≈200-300 A cm$^{-2}$ and peak output powers exceeding 1 W. These are strong figures of merit for devices emitting in this wavelength range. Using temperature-dependent measurements we explore the recombination processes occurring in the devices and find that, at room temperature, ≈24% of J$_{th}$ occurs through radiative recombination with the remaining 76% consistent with non-radiative thermally activated processes. While observations of non-pinning of the low-temperature integrated spontaneous emission above threshold suggests that a degree of inhomogeneity and localised states are present in the device, the strong reduction in differential efficiency at higher temperatures suggests that thermally activated carrier leakage-related processes are responsible for the relatively high levels of non-radiative recombination at room temperature and suggests a direct route for further improving device performance through improvements in conduction band offset (e.g. by incorporating dilute amounts of Al into the GaAs barriers). However, despite the presence of leakage in the investigated design, the low absolute threshold current densities along with the reduced temperature sensitivity of the lasing wavelength and high pulsed output powers indicate that type-II "W"-QWH lasers offer significant potential for lasers operating at optical communications wavelengths.

## Acknowledgements

The authors at the University of Surrey gratefully acknowledge EPSRC for funding this work under grant EP/N021037/1, in addition to SEPnet for funding a studentship for D. A. Duffy. This work was also funded in part by the German Science Foundation (DFG) through the Collaborative Research Centre CRC 1083 "Structure and Dynamics of Interior Interfaces".